\documentclass[epj]{svjour} 

\usepackage{graphicx}
\usepackage{subfigure}
\usepackage{epsfig}
\usepackage{amssymb}
\usepackage{mathrsfs}
\usepackage{calrsfs}
\usepackage{graphics}
\usepackage{txfonts}
\usepackage{pslatex}

\newcommand{\sqrtsnn}{\sqrt{s_{\mbox{\tiny{\it{NN}}}}}}
\newcommand{\sqrts}{\sqrt{s}}
\def\mean#1{\ensuremath{\left<#1\right>}}

\def\Lambdaqcd{\Lambda_{\ensuremath{\it QCD}}}

\newcommand{\jpsi}{J/\psi}

\newcommand{\ups}{\Upsilon}

\newcommand{\dNdeta}{dN_{\ensuremath{\it ch}}/d\eta|_{\eta=0}}

\newcommand{\gaga}{\gamma\,\gamma}
\newcommand{\gp}{\gamma\,p}
\newcommand{\gA}{\gamma\,A}

\begin{document}
\title{Low-$x$ QCD physics from RHIC and HERA to the LHC}
\author{David d'Enterria\inst{1}}
\institute{$^1$CERN, PH-EP, CH-1211 Geneva 23, Switzerland}


%
\date{Received: date / Revised version: date}
%
\abstract{We present a summary of the physics of gluon saturation and non-linear QCD evolution 
at small values of  parton momentum fraction $x$ in the proton and nucleus in the context of 
recent experimental results at HERA and RHIC. The rich physics potential of low-$x$ studies 
at the LHC, especially in the forward region, is discussed and some benchmark measurements 
in $pp$, $pA$ and $AA$ collisions are introduced.%
} 

\maketitle

\section{Introduction} 
\label{sec:intro}

\subsection{Parton structure and evolution}

\begin{figure}[htb]
\begin{center}
\epsfig{file=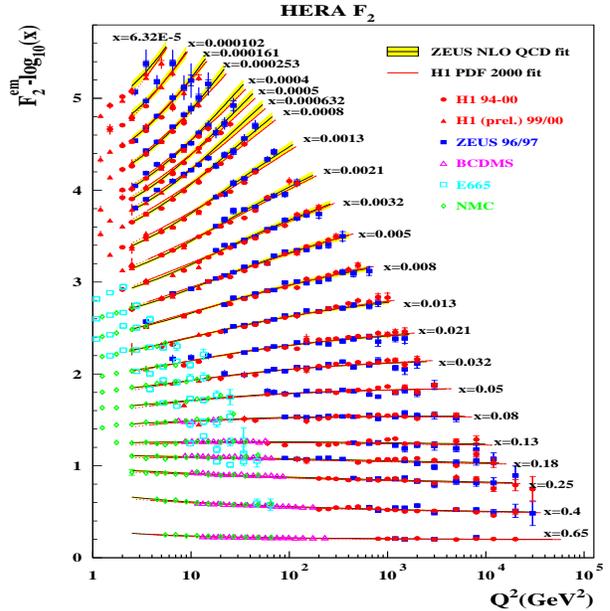,width=8.5cm,height=8.5cm}
\caption{$F_2(x,Q^2)$ measured in proton DIS at HERA ($\sqrt{s}$ = 320 GeV) and fixed-target 
($\sqrt{s}\approx$ 10-30 GeV) experiments
\label{fig:HERA_F2}}
\end{center}
\vspace{-.7cm}
\end{figure}

\begin{figure}[htb]
\begin{center}
\epsfig{file=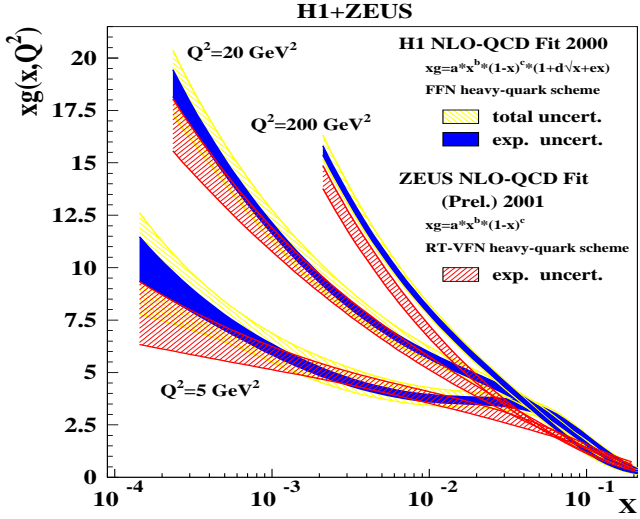, width=8.5cm,height=7.5cm}
\caption{Gluon distributions extracted at HERA (H1 and ZEUS) as a function 
of $x$ in three bins of $Q^2$~\protect\cite{hera_lhc05}
\label{fig:HERA_xG}}
\end{center}
\vspace{-.4cm}
\end{figure}

\begin{sloppypar}
The partonic structure of the proton (nucleus) can be probed with high precision in
deep inelastic scattering (DIS) electron-proton $ep$ (electron-nucleus, $eA$) collisions.
The inclusive DIS hadron cross section, $d^2\sigma/dx\,dQ^2$,  is a function of 
the virtuality $Q^2$ of the exchanged gauge boson  (i.e. its ``resolving power''), and 
the Bjorken-
$x$ fraction of the total nucleon momentum carried by the struck parton. 
The differential cross section 	for the neutral-current ($\gamma,Z$ exchange) process 
can be written in terms of the target structure functions as
\begin{equation}
\frac{d^2\sigma}{dx\,dQ^2} = \frac{2\pi\alpha^2}{x\,Q^4}\left[Y_{+}\cdot F_2\mp Y_{\_}\cdot x F_3-\vary^2\cdot F_L\right]\,,
\end{equation}
where $Y_{\pm}=1\pm(1-\vary)^2$ is related to the collision inelasticity $\vary$, and the 
structure functions $F_{2,3,L}(x,Q^2)$ describe the density of quarks and gluons in the hadron:
$F_2\propto e^2_q\;x\;\Sigma_i(q_i+\bar{q}_i)$, $xF_3\propto x\,\Sigma_i(q_i-\bar{q}_i)$, 
$F_L\propto \alpha_s\,xg$ ($xq_i,x\bar{q}_i$ and $xg$ are the corresponding parton distribution 
functions, PDF). $F_2$, the dominant contribution to the cross section over most of phase space, 
is seen to rise strongly for decreasing Bjorken-$x$ at HERA (Fig.~\ref{fig:HERA_F2}).
The growth in $F_2$ is well described by $F_2(x,Q^2)\propto x^{-\lambda(Q^2)}$, with 
$\lambda \approx$ 0.1 -- 0.3 logarithmically rising with $Q^2$~\cite{adloff01}. 
The $F_2$ scaling violations evident at small $x$ in Fig.~\ref{fig:HERA_F2} are indicative of 
the increasing gluon radiation from sea quarks. The $xg(x,Q^2)$ distribution itself can be indirectly 
determined (Fig.~\ref{fig:HERA_xG}) from the $F_2$ slope:
\begin{equation}
\frac{\partial F_2(x,Q^2)}{\partial \ln (Q^2)} \approx \frac{10\,\alpha_s(Q^2)}{27\pi}\,xg(x,Q^2)\,.
\label{eq:xG}
\end{equation}
Although the PDFs are non-perturbative objects obtained from fits to the DIS data, once measured 
at an input scale $Q_0^2\gtrsim$ 2 GeV$^2$ their value at any other $Q^2$ can be determined 
with the Dokshitzer-Gribov-Lipatov-Altarelli-Parisi (DGLAP) evolution equations which govern the 
probability of parton branchings (gluon splitting, $q,g-$strahlung) in QCD~\cite{dglap}.
\end{sloppypar}

\begin{sloppypar}
The DGLAP parton evolution, however, only takes into account the $Q^2$-dependence of the PDFs,
effectively summing leading powers of $\left[\alpha_s \ln(Q^2)\right]^{n}$ (``leading twist'') 
generated by parton cascades in a region of phase space where the gluons have strongly-ordered
transverse momenta towards the hard subcollision $Q^2\gg k_{nT}^2\gg \cdots \gg k_{1T}^2$. 
Such a resummation is appropriate when $\ln(Q^2)$ is much larger than $\ln(1/x)$. 
For decreasing $x$, the probability of emitting an extra gluon increases as
$\propto\alpha_s\ln(1/x)$. In this regime, the evolution of parton densities 
proceeds over a large rapidity region,  $\Delta y\sim \ln(1/x)$, and the finite transverse momenta 
of the partons become increasingly important. Here 
the full $k_T$ phase space of the gluons (including scattering of off-shell partons) 
has to be taken into account and not just the strongly-ordered DGLAP part.
Thus, the appropriate description of the parton distributions is in terms of {\it $k_T$-unintegrated} 
PDFs, $xg(x,Q^2)=\int^{Q^2} dk_T^2 G(x,k_T^2)$ with $G(x,k_T^2)\sim h(k_T^2)x^{-\lambda}$,
described by the Balitski-Fadin-Kuraev-Lipatov (BFKL) equation~\cite{bfkl} which governs parton 
evolution in $x$ at fixed $Q^2$. Hints of extra BFKL radiation have been recently found at HERA in 
the enhanced production of forward jets  compared to DGLAP expectations~\cite{hera_fwd_jets,marquet05}.
At large $Q^2$, a description resumming over both $\alpha_s \ln(Q^2)$ and  $\alpha_s\ln(1/x)$ 
is given by the Ciafaloni-Catani-Fiorani-Marchesini (CCFM) evolution equation~\cite{ccfm}.
\end{sloppypar}

\subsection{Parton saturation and non-linear evolution at low $x$}

\begin{figure}[htb]
\begin{center}
\epsfig{figure=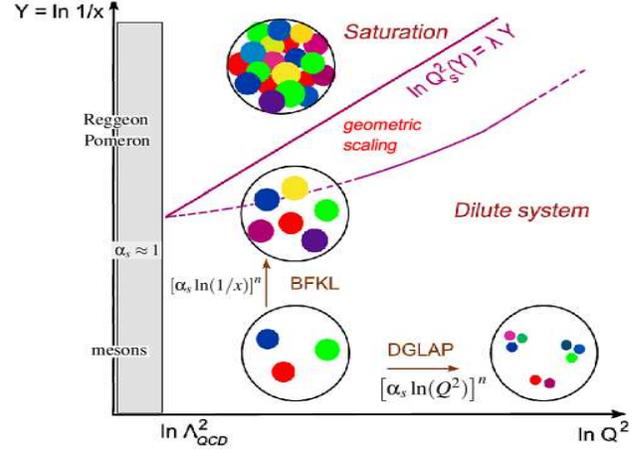, width=8.2cm,height=6.0cm}
\caption{QCD ``phase diagram'' in the $1/x,Q^2$ plane (each dot represents a parton with 
transverse area $\sim 1/Q^2$ and fraction $x$ of the hadron momentum). 
The different evolution regimes (DGLAP, BFKL, saturation) as well as the ``saturation
scale'' and ``geometric scaling'' curves between the dense and dilute domains are indicated. 
Adapted from~\protect\cite{iancu06}}
\label{fig:CGC_phase_diag}
\end{center}
\vspace{-.4cm}
\end{figure}

\begin{sloppypar}
As shown in Figure~\ref{fig:HERA_xG}, the gluon density rises very fast for decreasing $x$. 
Eventually, at some small enough value of $x$ ($\alpha_s\ln(1/x) \gg 1$) the number of gluons 
is so large that non-linear ($gg$ fusion) effects become important, taming the growth of the 
parton densities. In such a high-gluon density regime three things are expected to occur: 
(i) the standard DGLAP and BFKL {\it linear} equations should no longer be applicable since 
they only account for single parton branchings ($1\rightarrow$2 processes) 
but not for non-linear ($2\rightarrow$1) gluon recombinations; 
(ii) pQCD (collinear and $k_T$) factorization should break due to its (now invalid) 
assumption of {\it incoherent} parton scattering; and, as a result, 
(iii) standard pQCD calculations lead to a {\it violation of unitarity} 
even for $Q^2\gg \Lambdaqcd^2$. Figure~\ref{fig:CGC_phase_diag}  schematically depicts
the different parton evolution regimes as a function of $y=\ln(1/x)$ and $Q^2$. 
For small enough $x$ values and for virtualities below an energy-dependent ``saturation momentum'', 
$Q_s$, intrinsic to the {\it size} of the hadron, one expects to enter the regime of saturated
PDFs. Since $xg(x,Q^2)$ can be interpreted as the number of gluons with transverse 
area $r^2 \sim 1/Q^2$ in the hadron wavefunction, an increase of $Q^2$ effectively 
diminishes the `size' of each parton, partially compensating for the growth in their number 
(i.e. the higher $Q^2$ is, the smaller the $x$ at which saturation sets in). 
Saturation effects are, thus, expected to occur when the size occupied by the partons 
becomes similar to the size of the hadron, $\pi R^2$. In the case of nuclear targets with $A$ 
nucleons (i.e. with gluon density $xG=A\cdot xg$), this condition provides a definition 
for the saturation scale~\cite{GLR,MQ}:
\begin{equation}
Q_s^2(x)\simeq \alpha_s \frac{1}{\pi R^2}\,xG(x,Q^2)\sim A^{1/3}\,x^{-\lambda} \sim A^{1/3}(\sqrts)^{\lambda} \sim A^{1/3}e^{\lambda y},
\label{eq:Qs}
\end{equation}
with $\lambda\approx$ 0.25~\cite{kharzeev_kln}. Eq.~(\ref{eq:Qs}) tell us that $Q_s$ grows with 
the number of nucleons in the target and with the energy of the collision, $\sqrts$, or equivalently, 
the rapidity of the gluon $y=\ln(1/x)$. The nucleon number dependence implies that, at equivalent 
energies, saturation effects will be enhanced by factors as large as $A^{1/3}\approx$ 6 in heavy 
nuclear targets ($A$ = 208 for $Pb$) compared to protons.
In the last fifteen years, an effective field theory of QCD in the high-energy (high density, small $x$) 
limit has been developed - the Colour Glass Condensate (CGC)~\cite{cgc} - which describes the 
hadrons in terms of classical fields (saturated gluon wavefunctions) below the saturation scale $Q_s$. 
The saturation momentum $Q_s$ introduces a (semi-)hard scale, $Q_s\gg\Lambdaqcd$, 
which not only acts as an infrared cut-off to unitarize the cross sections but allows weak-coupling 
perturbative calculations ($\alpha_s(Q_s)\ll$1) in a strong $F_{\mu\nu}$ colour field background.
In the CGC framework, hadronic and nuclear collisions are seen as collisions of classical wavefunctions 
which ``resum'' all gluon  recombinations and multiple scatterings.
The quantum evolution in the CGC approach is given by the JIMWLK~\cite{jimwlk} non-linear equations
(or by their mean-field limit for $N_c\rightarrow\infty$, the Balitsky-Kovchegov equation~\cite{bk}) 
which reduce to the standard BFKL kernel at higher $x$ values.
\end{sloppypar}


\section{Parton saturation: Experimental studies}
\label{sec:signatures}

\begin{figure}[htb]
\begin{center}
\epsfig{file=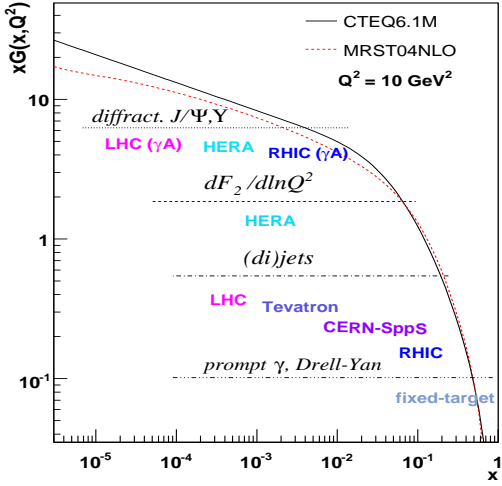,width=7cm,height=6.8cm}
\end{center}
\vspace{-0.4cm}
\caption{Examples of experimental measurements at various facilities providing information 
on the gluon PDF in the range $x\sim 10^{-5}-0.8$}
\label{fig:xG_processes}
\end{figure}

\begin{sloppypar}
The main  source of information on the PDFs is obtained from hard processes as they involve
outgoing particles directly coupled to the partonic scattering vertices. Figure~\ref{fig:xG_processes} 
summarizes the variety of measurements at different experimental facilities which are sensitive 
to the gluon density and their approximate $x$ coverage. $xG$ enters directly at LO in hadron-hadron 
collisions with (i) prompt photons, (ii) jets, and (iii) heavy-quarks in the final state, as well as in the 
(difficult) DIS measurement\footnote{$xG$ can also be (in)directly extracted from $F_2$ through 
the derivative in Eq.~(\ref{eq:xG}) as well as from the $F_2^{charm}$ data~\cite{hera_lhc_heavyQ}.} 
of (iv) the longitudinal structure function $F_L$. In addition, (iv) heavy vector mesons ($\jpsi,\ups$) 
from diffractive photoproduction processes\footnote{Diffractive $\gp$ ($\gA$) processes are 
characterized by a quasi-elastic interaction - mediated by a Pomeron or two gluons in a colour 
singlet state - in which the $p\;(A)$ remains intact (or in a low excited state) and separated 
by a rapidity gap from the rest of final-state particles.} are a valuable probe of the gluon density 
since their cross sections are proportional to the {\it square} of $xG$~\cite{ryskin95,teubner05}:
\begin{eqnarray}
\left .\frac{d\sigma_{\gp,A\rightarrow V\,p,A}}{dt}\right|_{t=0} =
\frac{\alpha_s^2\Gamma_{ee}}{3\alpha M_V^5}16\pi^3\left[xG(x,Q^2)\right]^2\,, \label{eq:diffract_qqbar_sigma}\\
\;\;\mbox{with }\;Q^2=M_V^2/4\;\;\mbox{ and }\;x=M_V^2/W_{\gp,A}^2.
\label{eq:diffract_qqbar_x}
\end{eqnarray}
The main  source of information on the {\it quark} densities is obtained from measurements of 
(i) the structure functions $F_{2,3}$ in lepton-hadron scattering, and 
(ii) lepton pair (Drell-Yan, DY) production in hadron-hadron collisions. 
In hadronic collisions, one commonly measures the perturbative probes 
at central rapidities ($y=0$) where $x=x_T=Q/\sqrts$, and $Q\sim p_T,M$  is the characteristic scale 
of the hard scattering. However, one can probe smaller $x_2$ values in the target by measuring 
the corresponding cross sections in the {\it forward} direction. Indeed, for a $2\rightarrow 2$ 
parton scattering the {\it minimum} momentum fraction probed in a process with a particle 
of momentum $p_T$ produced at pseudo-rapidity $\eta$ is~\cite{vogels_dAu}
\begin{equation}
x_{2}^{min} = \frac{x_T\,e^{-\eta}}{2-x_T\,e^{\eta}}\;\; \mbox{ where } \;\; x_T=2p_T/\sqrt{s}\,,
\label{eq:x2_min}
\end{equation}
i.e. $x_2^{min}$ decreases by a factor of $\sim$10 every 2 units of rapidity. Though Eq.~(\ref{eq:x2_min}) 
is a lower limit at the end of phase-space (in practice the $\mean{x_2}$ values in parton-parton
scatterings are at least 10 larger than $x_2^{min}$~\cite{vogels_dAu}), it provides the right 
estimate of the typical $x_2=(p_T/\sqrts)\,e^{-\eta}$ values reached in non-linear $2\rightarrow 1$ 
processes (in which the momentum is balanced by the gluon ``medium'') as described in parton 
saturation models~\cite{accardi04,dumitru05}.
\end{sloppypar}

\begin{figure}[htb]
\begin{center}
\epsfig{file=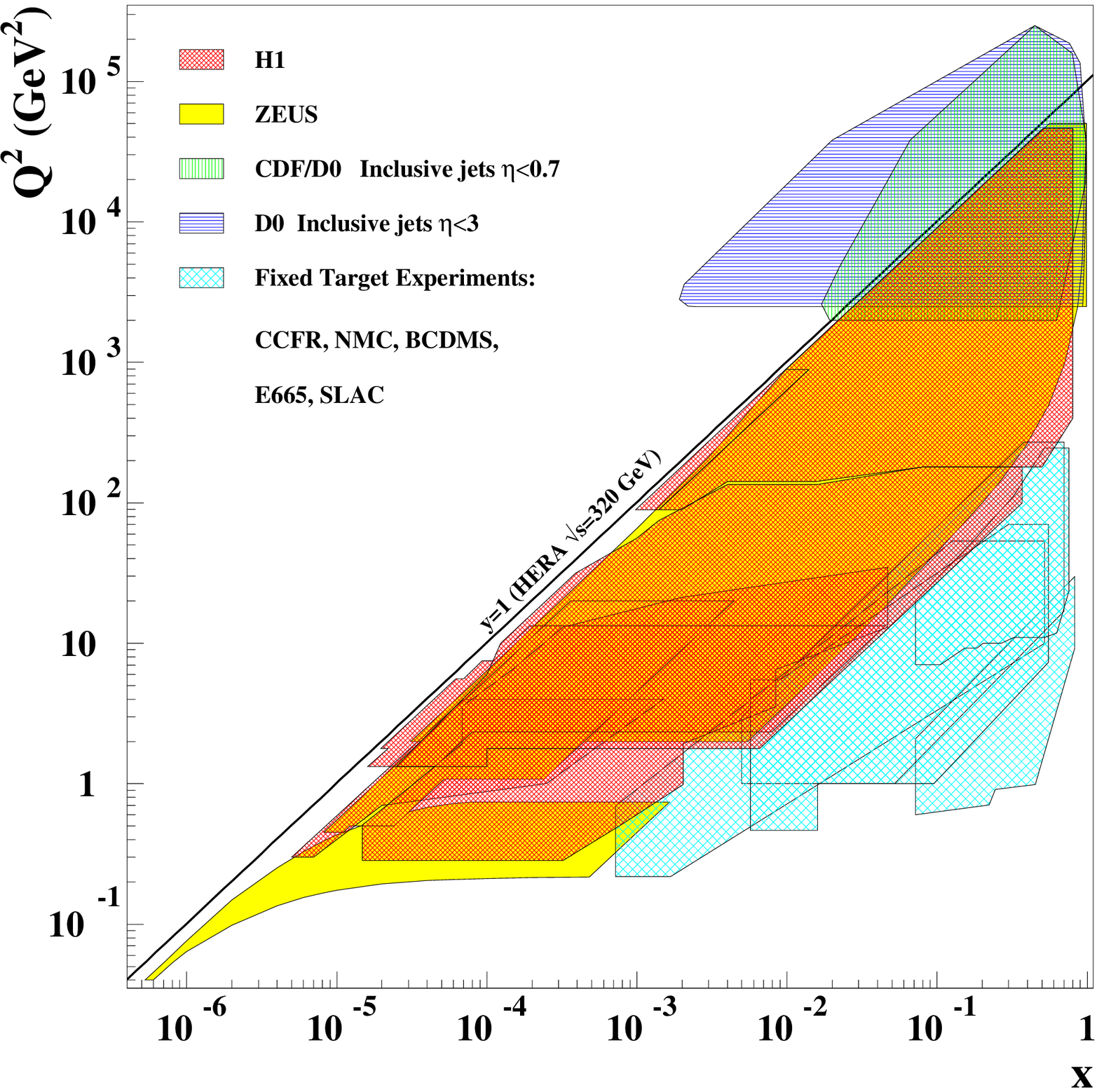,width=7.cm,height=5.cm}
\hspace*{-.3cm}
\epsfig{file=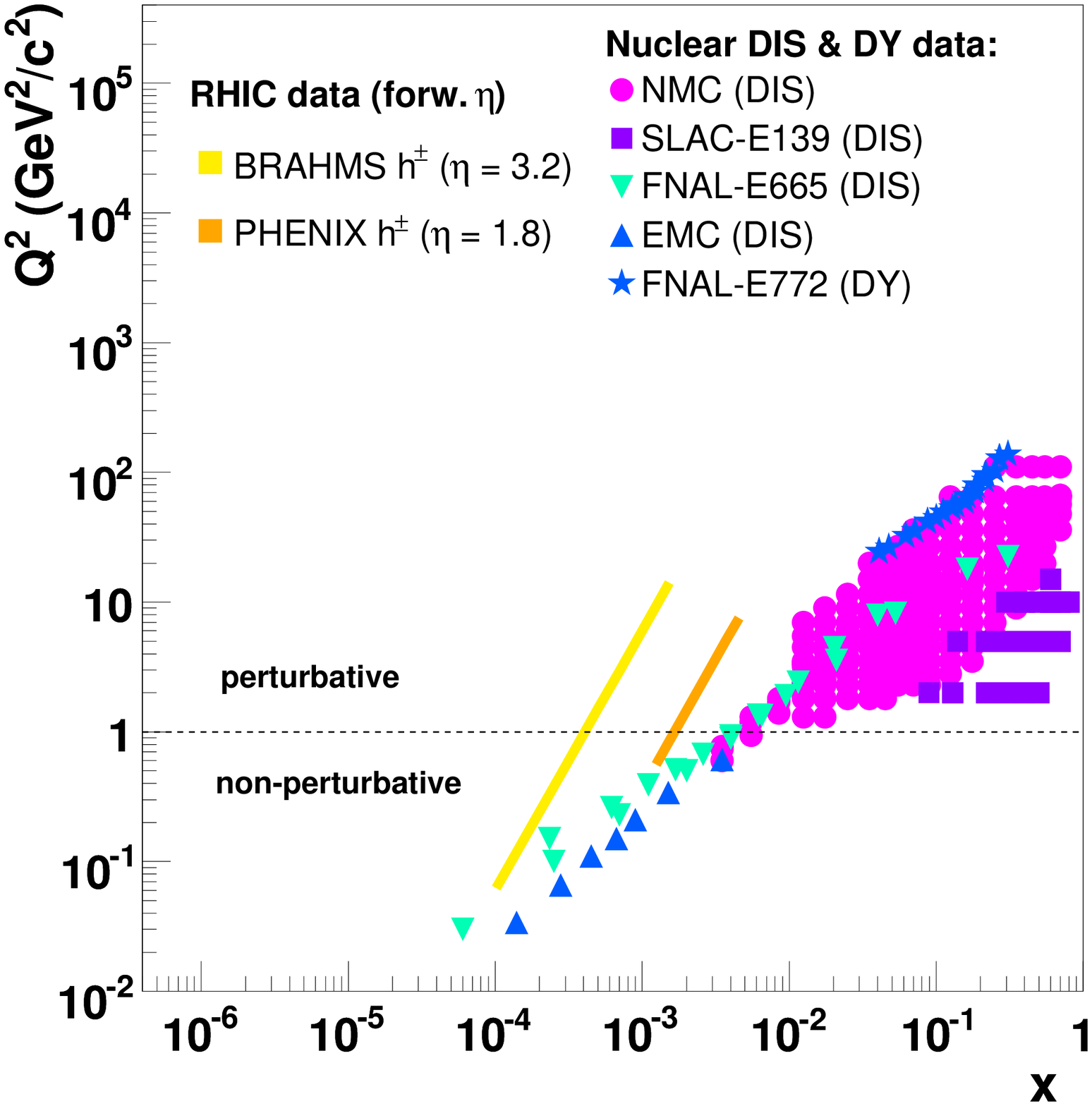,width=7.1cm,height=5.2cm}
\caption{Available measurements in the  $(x,Q^2)$ plane  used for the 
determination of the proton~\protect\cite{newman03} (top) and nuclear~\protect\cite{dde_qm04} (bottom) PDFs}
\label{fig:x_Q2_map}
\end{center}
\vspace{-0.7cm}
\end{figure}

\begin{figure}[htb]
\begin{center}
\epsfig{file=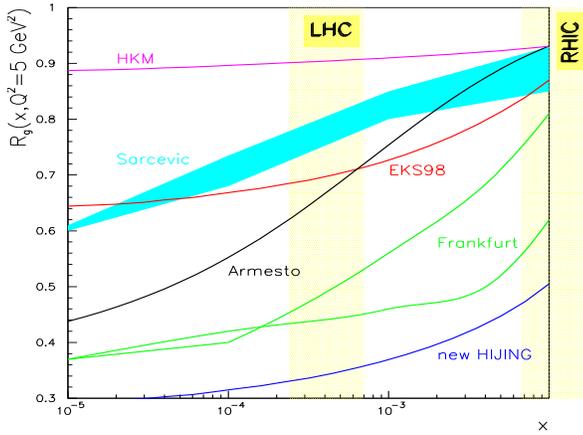,width=8.cm,height=6.5cm}
\caption{Ratios of the $Pb$ over proton gluon PDFs versus $x$ from different models
at $Q^2$ = 5 GeV$^2$. 
Figure taken from~\protect\cite{yellowrep_pdf}}
\label{fig:xG_Pb}
\end{center}
\vspace*{-.5cm}
\end{figure}

\begin{sloppypar}
Figure~\ref{fig:x_Q2_map} shows the kinematical map in $(x,Q^2)$ of the existing DIS, DY, direct 
$\gamma$ and jet data used in the PDF fits. Results from HERA and the Tevatron cover a substantial 
range of the proton structure ($10^{-4}\lesssim x\lesssim0.8$, 1 $\lesssim Q^2\lesssim$ 10$^{5}$ GeV$^2$) 
but the available measurements are much rarer  in the case of nuclear targets (basically 
limited to fixed-target studies, $10^{-2}\lesssim x\lesssim 0.8$ and
1 $\lesssim Q^2\lesssim$ 10$^{2}$ GeV$^2$). As a matter of fact, the nuclear parton distributions 
are basically unknown at low $x$ ($x<0.01$) where the only available measurements are
fixed-target data in the {\it non-perturbative} range ($Q^2<1$ GeV$^2$) dominated by
Regge dynamics rather than quark/gluon degrees of freedom. An example of the current
lack of knowledge of the nuclear densities at low $x$ is presented in Fig.~\ref{fig:xG_Pb}
where different available parametrizations of the ratio of $Pb$ to proton gluon distributions,
consistent with the available nDIS data at higher $x$, show differences as large as a factor 
of three at $x\sim 10^{-4}$~\cite{yellowrep_pdf,armesto_shadow}.
\end{sloppypar}

\subsection{HERA results}
\label{sec:hera}

\begin{figure}[htb]
\begin{center}
\hspace*{.5cm}
\centerline{\epsfig{file=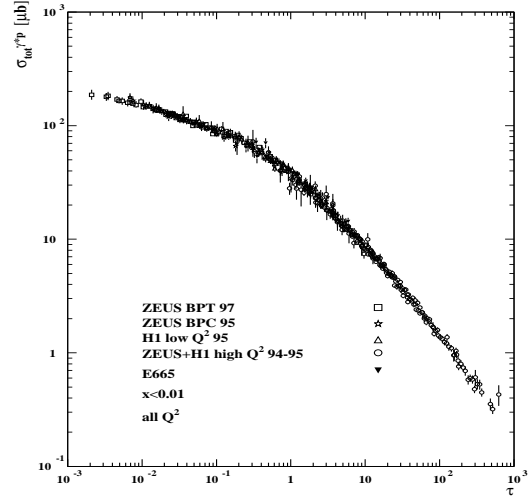,width=8.7cm,height=7.cm}}
\caption{Geometric scaling in the DIS $\gamma^\star p$ cross sections plotted 
versus $\tau=Q^2/Q_s^2$ in the range $x < 0.01$, 
0.045 $< Q^2 <$ 450 GeV$^2$~\protect\cite{golec_biernat_wusthoff}}
\label{fig:geom_scaling}
\end{center}
\vspace*{-.4cm}
\end{figure}

\begin{sloppypar}
Though the large majority of $ep$ DIS data collected during the HERA-I phase are
successfully reproduced by standard DGLAP predictions, 
more detailed and advanced experimental and theoretical results in the recent years have pointed 
to interesting hints of non-linear QCD effects in the data. Arguably, the strongest indication 
of such effects is given by the so-called ``geometric scaling'' property observed in 
inclusive $\sigma_{DIS}$ for $x<0.01$~\cite{golec_biernat_wusthoff}
as well as in various diffractive cross sections~\cite{forshaw,marquet06}. For inclusive 
DIS events, this feature manifests itself in a total cross section at small $x$ ($x<0.01$) which 
is only a function of  $\tau\!=\!Q^2/Q_s^2(x)$, instead of being a function 
of $x$ and $Q^2/Q_s^2$ separately (Fig.~\ref{fig:geom_scaling}). The saturation momentum
follows $Q_s(x)=Q_0(x/x_0)^{\lambda}$ with parameters $\lambda\sim$ 0.3, $Q_0$ = 1 GeV, 
and $x_0\sim$ 3$\cdot$10$^{-4}$. 
Interestingly, the scaling is valid up to very large values of $\tau$, well above the saturation scale, 
in an ``extended scaling'' region (see Fig.~\ref{fig:CGC_phase_diag}) where 
$Q_s^2<Q^2<Q_s^4/\Lambdaqcd^2$~\cite{ext_scal,iancu06}. 
The saturation formulation is suitable to describe not only inclusive DIS, but also inclusive diffraction 
$\gamma^\star p \rightarrow X\,p$. The very similar energy dependence of the inclusive diffractive 
and total cross sections in $\gamma^\star p$ collisions at a given $Q^2$ is easily explained 
in the Golec-Biernat W\"usthoff model~\cite{golec_biernat_wusthoff} but not in standard
collinear factorization. Furthermore, geometric scaling has been also found in different diffractive
DIS cross sections (inclusive, vector mesons, deeply-virtual Compton scattering DVCS)~\cite{forshaw,marquet06}. 
All the observed scalings are suggestive manifestations of the QCD saturation regime. 
Unfortunately, the values of $Q_s^2\sim$ 0.5 - 1.0 GeV at HERA lie in the transition 
region between the soft and hard sectors and, therefore, non-perturbative effects obscure the 
obtention of clearcut experimental signatures.
\end{sloppypar}

\subsection{RHIC results}
\label{sec:rhic}

\begin{sloppypar}
The expectation of enhanced parton saturation effects in the {\it nuclear} wave functions accelerated 
at ultra-relativistic energies, Eq.~(\ref{eq:Qs}), has been one of the primary physics motivations
for the heavy-ion programme at RHIC\footnote{At $y$ =0, the saturation scale  for a $Au$ nucleus 
at RHIC ($Q_s^2\sim$ 2 GeV) is larger than that of protons probed at HERA ($Q_s^2\sim$ 0.5 GeV).}~\cite{cgc}.
Furthermore, the properties of the high-density matter produced in the final-state of $AA$ interactions cannot be properly 
interpreted without having isolated first the influence of {\it initial state} modifications of the nuclear PDFs.
After five years of operation, two main experimental results at RHIC have been found consistent with CGC 
predictions: (i) the modest hadron multiplicities measured in $AuAu$ reactions, and (ii) the suppressed hadron 
yield observed at forward rapidities in $dAu$ collisions. [In addition, a recent analysis of the existing nuclear DIS 
$F_2$ data also confirms the existence of ``geometrical scaling'' for $x<$0.017~\cite{armesto04}.]
\end{sloppypar}

\begin{figure}[htb]
\begin{center}
\vspace{-0.6cm}
\epsfig{file=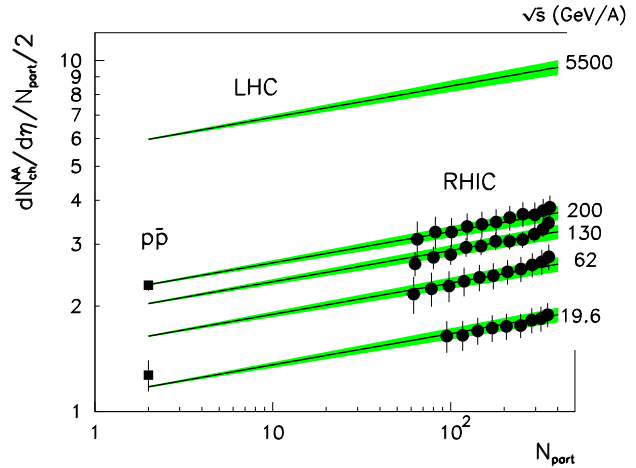,width=8.4cm}
\caption{Dependences on c.m. energy and centrality (given in terms of the number of nucleons participating in the collision, 
$N_{\rm part}$) of  $\dNdeta$  (normalized by $N_{\rm part}$):
PHOBOS $AuAu$ data~\protect\cite{phobos_wp} vs the predictions of the saturation approach~\protect\cite{armesto04}}
\label{fig:dNdeta}
\end{center}
\vspace{-.4cm}
\end{figure} 

\begin{sloppypar}
The bulk hadron multiplicities measured at mid-rapidity in central $AuAu$ at $\sqrtsnn$ = 200 GeV are 
$\dNdeta\approx$ 700, 
comparatively lower than the $\dNdeta\approx$ 1000 expectations of ``minijet'' dominated 
scenarios~\cite{hijing},  soft Regge models~\cite{dpm} (without accounting for strong 
shadowing effects~\cite{perco}), or extrapolations from an incoherent sum of proton-proton 
collisions~\cite{eskola_qm01}. On the other hand, CGC approaches~\cite{kharzeev_kln,armesto04} 
which effectively take into account a reduced number of scattering centers in the nuclear PDFs, 
$f_{a/A}(x,Q^2)<A\cdot f_{a/N}(x,Q^2)$ reproduce well not only the measured hadron 
multiplicities but - based on the general expression~(\ref{eq:Qs}) - also the centrality and c.m. 
energy dependences of the bulk $AA$ hadron production (Fig.~\ref{fig:dNdeta}).
\end{sloppypar}

\begin{figure}[htb]
\begin{center}
\includegraphics[width=8cm,height=5.5cm]{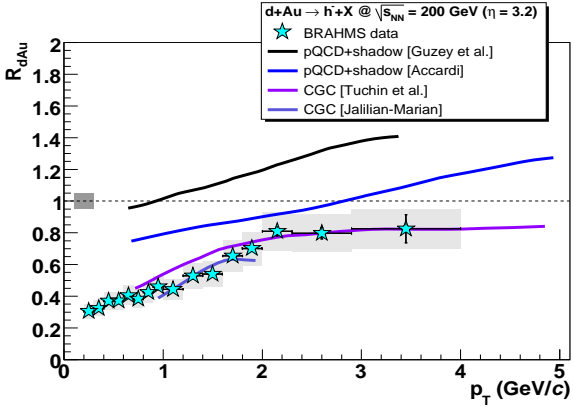}
\caption{Nuclear modification factor $R_{dAu}(p_T)$ for negative hadrons at $\eta$ =3.2 in 
$dAu$  at $\sqrtsnn$ = 200 GeV: BRAHMS data~\protect\cite{brahms_wp} compared
to leading-twist shadowing pQCD~\protect\cite{vogels_dAu,accardi04} and 
CGC~\protect\cite{tuchin04,jamal04} predictions}
\label{fig:RdA}
\end{center}
\vspace{-.4cm}
\end{figure} 

\begin{sloppypar}
The second manifestation of saturation-like effects in the RHIC data is the BRAHMS observation~\cite{brahms_wp} 
of suppressed yields of moderately high-$p_T$ hadrons ($p_T\approx 2 - 4 $ GeV/$c$) in $dAu$ relative
to $pp$ collisions at $\eta\approx$ 3.2 (Fig.~\ref{fig:RdA}). Hadron production at such small angles is 
sensitive to partons in the $Au$ nucleus with $x_2\approx$ $\mathscr{O}$(10$^{-3}$).
The observed nuclear modification factor, $R_{dAu}\approx$ 0.8, cannot be reproduced by pQCD
calculations that include standard {\it leading-twist} shadowing of the nuclear PDFs~\cite{vogels_dAu,accardi04}
but can be described by CGC approaches~\cite{tuchin04,jamal04} that parametrize the $Au$ nucleus as a saturated 
gluon wavefunction.
As in the HERA case, it is worth noting however that at RHIC energies the saturation scale is in the 
transition between the soft and hard regimes ($Q_s^2\approx$ 2 GeV$^2$) and the results consistent 
with the CGC predictions are in a kinematic range with relatively low momentum scales ($\mean{p_T}\sim$ 0.5 GeV
for the hadron multiplicities and  $\mean{p_T}\sim$ 2.5 GeV for forward inclusive hadron production)
where non-perturbative effects can blur a simple interpretation based on partonic degrees of freedom alone. 
\end{sloppypar}

\section{Low-$x$ QCD at the LHC}
\label{sec:lhc}

\begin{figure}[htb]
\begin{center}
\centerline{\epsfig{file=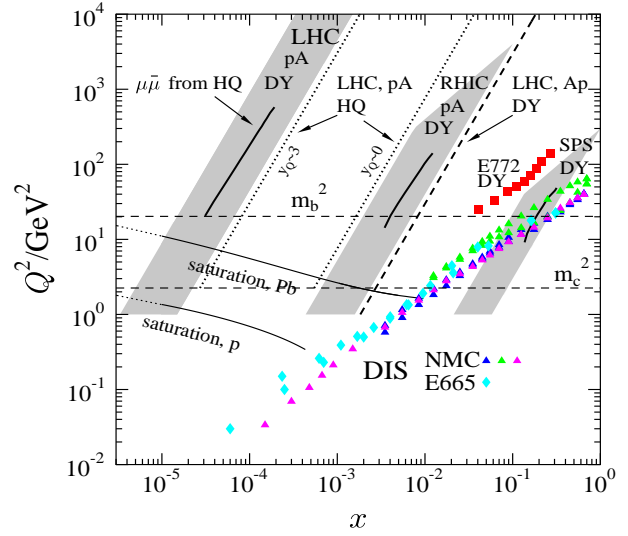,width=8.cm,height=7.cm}}
\caption{Kinematical $(x,Q^2)$ range probed 
at various rapidities $y$ and c.m. energies in $\sqrtsnn$ = 8.8 TeV $pA$ collisions at the LHC~\protect\cite{yellowrep_pdf}}
\label{fig:xQ2_map_pA_LHC}
\end{center}
\vspace*{-.5cm}
\end{figure}

\begin{sloppypar}
The Large Hadron Collider (LHC) at CERN will provide $pp$, $pA$ and $AA$ collisions 
at $\sqrtsnn$ = 14, 8.8 and 5.5 TeV respectively with luminosities $\mathscr{L}\sim$  10$^{34}$, 
10$^{29}$ and 5$\cdot$10$^{26}$ cm$^{-2}$ s$^{-1}$.
Such large c.m. energies and luminosities will allow detailed QCD studies at unprecedented 
low $x$ values thanks to the copious production of hard probes (jets, quarkonia, heavy-quarks,
prompt $\gamma$, Drell-Yan pairs, etc.).  
The advance in the study of low-$x$ QCD phenomena will be specially substantial for nuclear 
systems since the saturation momentum, Eq.~(\ref{eq:Qs}),  $Q_s^2 \approx$ 5 -- 10 GeV$^2$, 
will be in the perturbative range~\cite{kharzeev_kln}, and the relevant $x$ values, 
Eq.~(\ref{eq:x2_min}), will be 30--70 times lower than $AA$ and $pA$ reactions at RHIC: 
$x \approx 10^{-3}(10^{-5})$ at central (forward) rapidities for processes with 
$Q^2\sim$10 GeV$^2$ (Fig.~\ref{fig:xQ2_map_pA_LHC}).
\end{sloppypar}

\subsection{The LHC experiments}
\label{sec:cms}

\begin{sloppypar}
The four LHC experiments - i.e. the two large general-purpose and high-luminosity ATLAS and CMS 
detector systems as well as the heavy-ion-dedicated ALICE and the heavy-flavour-oriented LHCb 
experiments - have all detection capabilities in the forward direction very well adapted for the study 
of low-$x$ QCD phenomena with hard processes in collisions with proton and ion beams:
\begin{description}
\item (i) Both CMS and ATLAS feature hadronic calorimeters in the range 
3$<|\eta|<$5 which allow them to measure jet cross-sections at very forward rapidities.
Both experiments feature also zero-degree calorimeters (ZDC, $|\eta|\gtrsim$ 8.5 for neutrals), 
which are a basic tool for neutron-tagging ``ultra-peripheral'' $PbPb$ photoproduction interactions. 
CMS has an additional electromagnetic/hadronic calorimeter (CASTOR, 5.3$<|\eta|<$6.7)
and shares the interaction point with the TOTEM experiment providing two extra 
trackers at very forward rapidities (T1, 3.1 $<|\eta|<$ 4.7, and T2, 5.5$<|\eta|<$6.6) 
well-suited for DY measurements. 
\item (ii) The ALICE forward muon spectrometer at $2.5<\eta < 4$, can reconstruct
$\jpsi$ and $\ups$ (as well as $Z$) in the dimuon channel, as well as statistically measure 
single inclusive heavy-quark production via semileptonic (muon) decays. ALICE counts 
also on ZDCs in both sides of the interaction point (IP) for forward neutron triggering 
of $PbPb$ photoproduction processes.
\item (iii) LHCb is a single-arm spectrometer covering rapidities $1.8<\eta<4.9$,  with 
very good particle identification capabilities designed to accurately reconstruct $b$ and $c$ 
mesons. The detector is also well-suited to measure jets, $Q\bar{Q}$ and Z$\rightarrow\mu\mu$ 
production in the forward hemisphere.
\end{description}
\end{sloppypar}

\subsection{Low-$x$ QCD measurements at LHC}
\label{sec:cms_lowx}
Measurement at the LHC forward rapidities of any of the processes shown in Fig.~\ref{fig:xG_processes}
provides an excellent means to look for signatures of high gluon density phenomena at low $x$. Four
representative measurements are discussed in the last section of the paper.

\subsubsection*{$\bullet$ Case study I: Forward (di)jets ($pp$, $pA$, $AA$)}
\begin{figure}[htb]
\begin{center}
\centerline{\epsfig{file=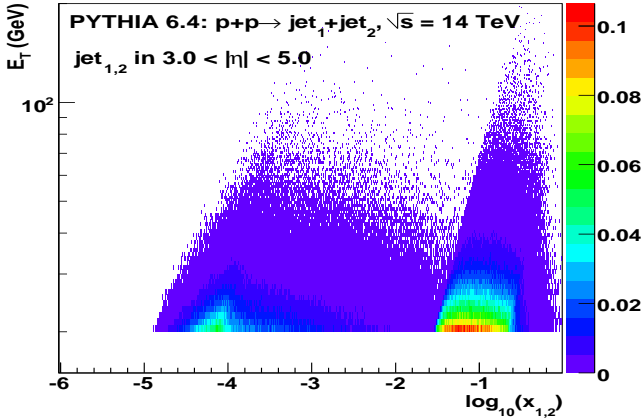,width=8.5cm,height=5.8cm}}
\caption{log($x_{1,2}$) distribution of two partons colliding in $pp$ collisions at $\sqrts$ = 14 TeV 
and producing at least one jet within ATLAS/CMS forward calorimeters acceptances as determined with PYTHIA}
\label{fig:x1x2_hf_jets}
\end{center}
\vspace*{-.8cm}
\end{figure}

\begin{sloppypar}
The jet measurements in $\bar{p}p$ collisions at Tevatron energies have provide valuable
information on the proton PDFs (see Fig.~\ref{fig:x_Q2_map}, top). According to Eq.~(\ref{eq:x2_min}), 
the measurement of jets with  $p_T\approx$ 20 - 200 GeV/$c$ in $pp$ collisions at 14 TeV 
in the ATLAS or CMS forward calorimeters (3$<|\eta|<$5) allows one to probe the PDFs
at $x$ values as low as $x_{2}\approx 10^{-4}-10^{-5}$. Fig.~\ref{fig:x1x2_hf_jets} 
shows the actual log($x_{1,2}$) distribution of two partons scattering at LHC and 
producing at least one forward jet as obtained with  PYTHIA 6.403~\cite{pythia6.4}. As expected 
in forward scattering, the collision is very asymmetric with $x_2$ ($x_1$) peaked at $\sim 10^{-4}$ 
($\sim 10^{-1}$) and thus provides direct information on the low-$x$ parton densities. 
Not only the single inclusive cross-section but the forward-backward dijet production, 
``M\"uller-Navelet jets'', is a particularly sensitive measure of BFKL~\cite{mueller_navelet} 
as well as non-linear~\cite{marquet05} parton evolutions. In the presence of low-$x$ 
saturation effects, the M\"uller-Navelet cross section for two jets separated by $\Delta\eta\sim$ 9 
(and, thus, measurable in each one of the forward calorimeters) is expected to be suppressed 
by a factor of $\sim$2 compared to BFKL predictions~\cite{marquet05}.  A study is 
underway to determine the feasibility of both forward (di)jet measurements in 
CMS~\cite{loi_cms_totem}.
\end{sloppypar}

\subsubsection*{$\bullet$ Case study II: Forward heavy-quarks ($pp$, $pA$, $AA$)}
\label{sec:heavyQ}
\begin{figure}[htb]
\begin{center}
\hspace*{.3cm}
\centerline{\epsfig{file=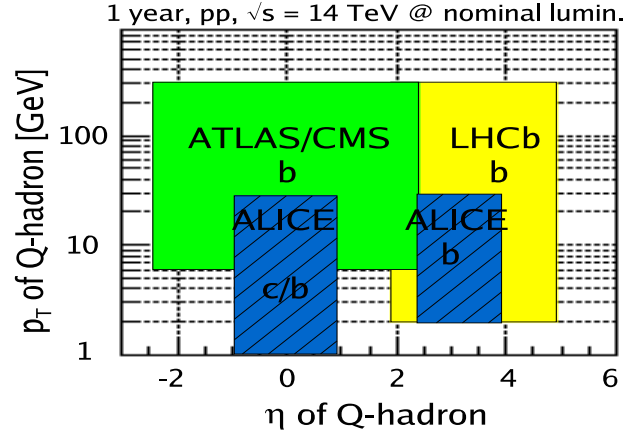,width=8.5cm,height=5.9cm}}
\caption{Acceptances in $(\eta,p_T)$ for open charm and bottom hadrons 
in the four LHC experiments for 1-year nominal luminosity~\cite{hera_lhc_heavyQ}}
\label{fig:heavyQ_accept_LHC}
\end{center}
\vspace*{-.8cm}
\end{figure}

Studies of small-$x$ effects on heavy flavour production at the LHC in two different 
approaches, based on collinear and $k_T$ factorization, including non-linear terms 
in the parton evolution, lead to two different predictions (enhancement vs. 
suppression) for the measured $c$ and $b$ cross-sections~\cite{hera_lhc_heavyQ}. 
The possibility of ALICE and LHCb to reconstruct $D$ and $B$ mesons in a large 
rapidity range (Fig.~\ref{fig:heavyQ_accept_LHC}) will put stringent constraints 
on the gluon structure and evolution at low-$x$. In the case of ALICE, the heavy-Q 
$pp$ studies will have a natural extension in $AA$ and $pA$ collisions~\cite{dainese_hp06}, 
providing a precise probe of non-linear effects in the nuclear wave-function.

\subsubsection*{$\bullet$ Case study III: $Q\bar{Q}$ photoproduction (electromagnetic $AA$)}
\label{sec:upc_qqbar}
\begin{sloppypar}
High-energy diffractive photoproduction of heavy vector mesons ($\jpsi,\ups$) proceeds through 
colourless two-gluon exchange (which subsequently couples to $\gamma\rightarrow Q\bar{Q}$) 
and is thus a sensitive probe of the gluon densities at small $x$, see Eq.~(\ref{eq:diffract_qqbar_sigma}).
Ultra-peripheral interactions (UPCs) of high-energy heavy ions generate strong electromagnetic fields 
which help constrain the low-$x$ behaviour of $xG$ via quarkonia~\cite{dde_qm05}, or other hard 
probes~\cite{strikman05}, produced in $\gamma$-nucleus collisions. 
Lead beams at 2.75 TeV have Lorentz factors $\gamma$ = 2930 leading to maximum (equivalent) 
photon energies $\omega_{\ensuremath{\it max}}\approx \gamma/R\sim$ 100 GeV,
and c.m. energies 
$W_{\gaga}^{\ensuremath{\it max}}\approx$ 160 GeV and $W^{\ensuremath{\it max}}_{\gA}\approx$ 1 TeV. 
From Eq.~(\ref{eq:diffract_qqbar_x}), the $x$ values probed in $\gA\rightarrow\jpsi \;A$ processes 
at $y$ = 2 can be as low as $x\sim 10^{-5}$. 
The ALICE, ATLAS and CMS experiment can measure $\jpsi,\ups\rightarrow e^+e^-,\mu^+\mu^-$ 
produced in electromagnetic $PbPb$ collisions tagged with neutrons detected in the ZDCs
(as done at RHIC~\cite{dde_qm05}). Figure 13 shows the expected dimuon invariant mass distributions 
around the $\ups$ mass predicted by {\sc starlight}~\cite{starlight} within the CMS central acceptance 
($\eta<$2.5) for an integrated $PbPb$ luminosity of 0.5 nb$^{-1}$~\cite{dde_lowx}. 
\end{sloppypar}

\begin{figure}[htb]
\begin{center}
\epsfig{file=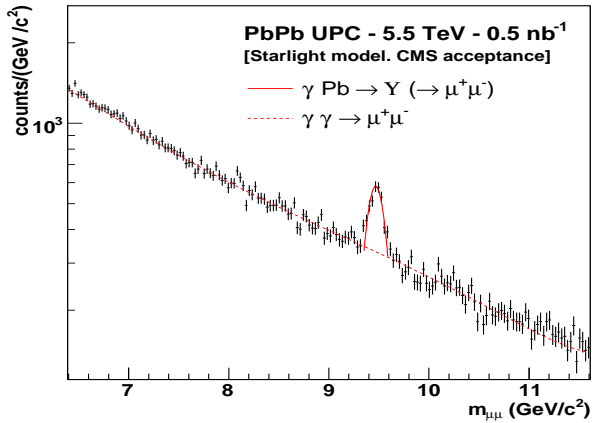,width=8.cm,height=6.cm}
\caption{Expected $\mu^+\mu^-$ invariant mass from $\gamma\,Pb\rightarrow \Upsilon\,Pb^\star\rightarrow \mu^{+}\mu^{-}\,Pb^\star$ 
and $\gaga\rightarrow \mu^+\mu^-$ processes predicted by {\it Starlight}~\protect\cite{starlight} 
for UPC $PbPb$ collisions at $\sqrtsnn$ = 5.5 TeV in the CMS acceptance}
\end{center}
\label{fig:upc_lhc_starlight}
\end{figure}

\subsubsection*{$\bullet$ Case study IV: Forward Drell-Yan pairs ($pp$, $pA$, $AA$)}
\begin{sloppypar}
High-mass Drell-Yan pair production at the very forward rapidities covered by LHCb and by the 
CMS CASTOR and TOTEM T2  detectors can probe the parton densities down to $x\sim 10^{-6}$
at much higher virtualities $M^2$ than those accessible in other measurements discussed here.
A study is currently underway in CMS~\cite{loi_cms_totem} to combine the CASTOR 
electromagnetic energy measurement together with the good position resolution of T2 for charged 
tracks, to trigger on and reconstruct the $e^+e^-$ invariant mass in $pp$ collisions at 14 TeV,
and scrutinize $xg$ in the $M^2$ and $x$ plane.
\end{sloppypar}

\section{Conclusion}
\label{sec:conclusion}

\begin{sloppypar}
We have reviewed the physics of non-linear QCD and high gluon densities at small fractional momenta 
$x$, with emphasis on the existing data at HERA (proton) and RHIC (nucleus) which support the 
existence of a parton saturation regime as described e.g. in the framework of the Colour Glass Condensate
effective field theory. The future perspectives at the LHC have been presented, including the promising 
capabilities of the forward detectors of the ALICE, ATLAS, CMS and LHCb experiments to study the 
parton densities down to $x\sim 10^{-6}$ with various hard probes (jets, quarkonia, heavy-quarks, 
Drell-Yan). The programme of investigating the dynamics of low-$x$ QCD is not only appealing in 
its own right but it is an essential prerequisite for predicting a large variety of hadron-, photon- and 
neutrino- scattering cross sections at very high energies.
\end{sloppypar}

\section*{Acknowledgments}
The author thanks A. de Roeck and H. Jung as well as A. Dobado and F. J. Llanes Estrada 
for their invitation to present this overview talk in the 2nd HERA-LHC Workshop and in 
the QNP'06 Int. Conf.
This work is supported by the 6th EU Framework Programme contract MEIF-CT-2005-025073.


\end{document}